\documentclass[aps,pre,superscriptaddress,twocolumn]{revtex4-1}

\usepackage{eurosym,amsmath,amssymb}
\usepackage{bm}
\usepackage{epsfig}
\usepackage{graphicx}
\usepackage[usenames,dvipsnames]{color}
\usepackage[colorlinks]{hyperref}
\usepackage[normalem]{ulem}
\hypersetup{colorlinks,citecolor=blue,linkcolor=blue,urlcolor=blue}  

\begin{document}

\title{Chaos in two-dimensional Kepler problem with spin-orbit coupling}
\author{V.A. Stephanovich}
\affiliation{Institute of Physics, Opole University, Opole, 45-052, Poland}
\author{E.Ya. Sherman}
\affiliation{Department of Physical Chemistry, University of the Basque Country UPV/EHU,
48080 Bilbao, Spain}
\affiliation{IKERBASQUE Basque Foundation for Science, Bilbao, Spain}
\date{\today}

\begin{abstract}
We consider classical two-dimensional Kepler system with spin-orbit coupling and show
that at a sufficiently strong coupling it demonstrates a chaotic behavior.
The chaos emerges since the spin-orbit coupling 
reduces the number of the integrals of motion as compared to the number of the degrees of freedom. 
This reduction is manifested in the equations of motion as the emergence of 
the anomalous velocity determined by the spin orientation.
By using analytical and numerical arguments, we
demonstrate that the chaotic behavior, being driven by this anomalous term,
is related to the system energy dependence on the initial spin orientation. 
We observe the critical dependence of the dynamics on the initial conditions, 
where system can enter and exit a stability domain
by very small changes in the initial spin orientation. 
Thus, this system can demonstrate a reentrant order-from-disorder transition driven 
by very small variations in the initial conditions.
\end{abstract}

\maketitle

\section{Introduction}

The emergence of a chaotic behavior is one of the most intriguing features
of dynamical systems \cite{Gutzwiller,Reichl,Haake,Stockmann,Casati}. {For instance, simple
dynamical systems with an energy-dependent separatrix (boundary
between qualitatively different trajectories in their phase space), demonstrate
chaos, e.g., being driven by a periodic external field.} 

Dynamical systems with spin-orbit coupling (SOC), ranging from 
semiconductors \cite{Dyakonov08} to ultracold atomic
matter \cite{spielman2013,zhaih2012}, 
became a topic of a great interest due to rich variety of physical effects
observable there. Here we propose and study a
class of nonstandard  classical \textit{conservative} dynamical systems, 
which can be characterized as two-dimensional (2D) spin-orbit coupled Kepler systems.
We show that inclusion of a new internal degree of freedom, 
namely, the particle spin, coupled to its momentum,
leads to the chaos emergence. 
 
Without SOC, the properties of the above systems are well-known in 
classical \cite{Landau} and quantum \cite{Portnoi} realization and no 
chaotic behavior is expected there except a quantum chaos in an applied magnetic field 
acting at the orbital electron motion \cite{Friedrich}.
{Three integrals of motion: the energy, the angular 
momentum, and the Runge-Lenz vector fully determine the dynamics of the system 
and assure its stability against chaos.} The SOC introduces new degrees of freedom and lowers the system symmetry to 
the existence of only two integrals of motion. This symmetry reduction can be seen as 
a SOC-generated spin-dependent contribution to the velocity proportional to the
SOC constant.  As a result, the system loses integrability due to the spin back action on the orbital
motion and can demonstrate a chaotic behavior close to the separatrix which determines the boundary
between a finite and a delocalized motion.

An experimentally realizable example of such a system is given by 2D
excitons in semiconductor structures \cite{Durnev,Vishnevsky,High}.
We shall consider the classical limit of orbital motion corresponding to the highly excited 
Rydberg states of these excitons. {Recently, it has been} experimentally established  
that lowering of the system symmetry from the vacuum SO(4) to the discreet
symmetry of a host crystal for three-dimensional excitons leads to a chaos even in the SOC 
absence \cite{Assmann2016,Ostrovskaya2016,Schweiner2017}. Using only classical arguments, here we prove 
that the spin-orbit coupling, being a lower-symmetry contribution to the 
system dynamics, can induce the chaotic motion.  

{Two aspects of SOC-related randomness have recently been studied} in two-dimensional 
harmonic potentials, where particle's motion is always finite.
Larson \textit{et al.} \cite{Larson} studied thermalization
in cold atomic gases in the presence of anisotropic SOC, 
while Marchukov \textit{et al.} \cite{Marchukov}
examined the spectral properties in terms of the ensembles emerging in the random matrix theory. 
In addition, a chaos-like behavior in driven SOC systems with a strong confinement {has been}
considered in Refs. \cite{Khomitsky} and \cite{Chotorlishvili}. 
Conservative two-dimensional systems with interaction potentials vanishing at large distances, being augmented by the SOC, 
provide nontrivial examples of a classical chaotic behavior, qualitatively different from above settings. 
Motivated by possible transitions to the nontrivial chaotic behavior, here we study the effects of weak and strong SOC on the 
dynamics  of a Kepler-like system, prove that such transitions indeed occur, and analyze different regimes of the chaotic motion. 

This paper is organized as follows. In Section II we introduce general equations and integrals of motion for the Kepler system 
with spin-orbit coupling. In Section III we perform perturbative analysis of the trajectories at a weak SOC. In Section IV
the chaotic behavior at a sufficiently strong SOC will be presented and analyzed. Conclusions and relation to possible 
experiments will be given in Section V.

\section{Equations and integrals of motion}

\begin{figure}[t]
\begin{center}
\includegraphics[width=0.7\columnwidth]{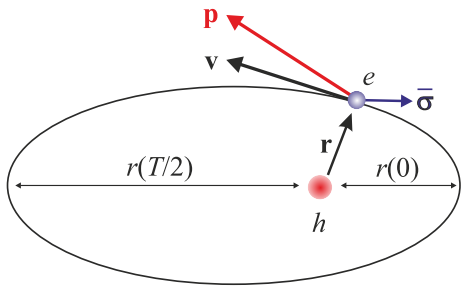}
\end{center}
\caption{ Single-period trajectory with $\bar{\sigma}_{x}=1$ 
at a weak SOC. Notations correspond to the text. Electron $(e)$ starts to move at
the distance $r(0)$ from the fixed hole $(h)$ and at $t=T/2,$ where $T$ is the rotation period, reaches the 
distance $r(T/2).$}
\label{fig:trajectory}
\end{figure}

We take the minimal Hamiltonian describing the 2D Kepler 
problem with SOC in the form $H=H_{0}+H_{\mathrm{so}}.$
The spin-independent part is: 
\begin{equation}
H_{0}=\frac{p^{2}}{2m}-\frac{e^{2}}{r},  \label{ham1}
\end{equation}%
where $\mathbf{r}=(x,y)$ is the electron coordinate,   $\mathbf{p}$ is the 
electron momentum (see Fig. \ref{fig:trajectory}), 
$m$ is its mass, and $e$ is the effective charge including the dielectric
constant of a host crystal. We take the SOC term in the Rashba form \cite{Bychkov} 
with rotational symmetry of the spectrum: 
\begin{equation}
H_{\mathrm{so}}=\frac{\alpha }{\hbar }\left( p_{x}{\sigma }_{y}-p_{y}{\sigma}_{x}\right),  
\label{hso}
\end{equation}%
with $\alpha $ being the coupling constant \cite{alpha} and ${\sigma }_{i}$
are the Pauli matrices corresponding to spin $1/2$. 
{This type of coupling appears as a result of either spatial 
inversion asymmetry in solids \cite{Bychkov,Fabian} or can 
be generated optically in cold atomic gases \cite{spielman2013,zhaih2012}.}

Although here we concentrate on the classical dynamics, it is instructive 
to make a connection with the quantum approach. Namely, we consider a system with 
two-component wavefunction in the form:
\begin{equation}
{\bm\psi}\left({\mathbf{r}},t\right)={\varphi}\left({\mathbf{r}},t\right)
\left[
\begin{array}{ll}
\cos\left(\theta/2\right)e^{i\phi}\\
\sin\left(\theta/2\right)
\end{array}
\right].
\end{equation}
Here ${\varphi}({\mathbf{r}},t)$ is {the (spatial and temporal dependent) 
wavepacket-like envelope function} and time-dependent 
angles $\theta$ and $\phi$ determine spin components as:
\begin{equation}
{\sigma}_{z}=\cos\theta,\quad {\sigma}_{x}=\sin\theta\,\cos\phi, 
\quad {\sigma}_{y} = -\sin\theta\,\sin\phi.
\end{equation}
Subsequently we assume the classical limit for orbital motion, while 
the spin remains quantum.
To be specific, here we characterize the orbital motion by the classical 
coordinate $\mathbf{r}$ and momentum $\mathbf{p}$, whereas \textit{quantum} 
spin defines the momentum-dependent precession \cite{Larson,Fabian} and related to 
this precession effects, as described below.  

The characteristic feature of Hamiltonian \eqref{hso} is the anomalous 
spin-dependent velocity \cite{Adams} (see Fig. \ref{fig:trajectory}) presented 
in the commutator form $\mathbf{v}_{\mathrm{so}}\equiv i[H_{\mathrm{so}},%
\mathbf{r}]/\hbar=\alpha\left( {\sigma }_{y},-{\sigma }_{x}\right)/\hbar$. 
Two other SOC characteristics \cite{Fabian} of our interest are the spin precession 
with the rate $2p\alpha/\hbar^{2}$ and
the corresponding length $l_{\mathrm{so}}=\hbar^{2}/m\alpha $, necessary for electron
to essentially rotate the spin. 

To formulate the classical equations of motion, we observe from Eq. \eqref{hso} that the above
anomalous velocity is the derivative $\mathbf{v}_{\mathrm{so}}=\partial H_{%
\mathrm{so}}/\partial \mathbf{p}$. For Hamiltonian function $H=H_{0}+H_{\mathrm{so}}$ 
these equations can be
obtained by substituting the spin components in Eq. \eqref{hso} by their
expectation values $\bar{\sigma}_{i}$ such that $\bar{\sigma}_{x}^{2}+\bar{%
\sigma}_{y}^{2}+\bar{\sigma}_{z}^{2}=1$. This yields 
\begin{equation}
\dot{\mathbf{p}}=-e^{2}\frac{\mathbf{r}}{r^{3}},\quad \dot{\mathbf{r}}=\frac{\mathbf{p}}{m}-
\frac{\alpha}{\hbar}\left[ \mathbf{z}\times \bar{\bm\sigma }\right] ,  \label{dpdt}
\end{equation}%
where $\mathbf{z}=(0,0,1)$ is the unit vector perpendicular to $(xy)$ plane,
where the motion occurs. The equations for spin components corresponding to the 
precession dependent on the particle momentum read: 
\begin{eqnarray}
\dot{\bar{\sigma}}_{x}&=&2\frac{\alpha\,m}{\hbar^{2}} \left( \dot{x}-\frac{\alpha}{\hbar}\bar{\sigma}_{y}\right) \bar{\sigma}_{z},\notag \\
\dot{\bar{\sigma}}_{y}&=&-2\frac{\alpha\,m}{\hbar^{2}}\left(\dot{y}+\frac{\alpha}{\hbar}\bar{\sigma}_{x}\right) \bar{\sigma}_{z},  \notag \\
\dot{\bar{\sigma}}_{z}&=&-2\frac{\alpha\,m}{\hbar^{2}}\left( \dot{x}\bar{\sigma}_{x}+\dot{y}\bar{\sigma}_{y}\right).  \label{dsdt}
\end{eqnarray}%
Equations \eqref{dpdt} and \eqref{dsdt} permit the complete classical
analysis of particle position and spin components time dependence \cite{Bi}.
Their direct iterative and exact numerical solutions will be used for the description
of the trajectories.

Integrals of motion in this system are the energy $E,$ the $z-$component of
total angular momentum $\hbar\left(L+\bar{\sigma}_{z}/2\right)\equiv p_{y}x-p_{x}y+\hbar\bar{\sigma}%
_{z}/2$ and the length of the spin vector $\bar{\sigma}^{2}=1.$ The above
conservation laws can be verified by direct calculation. The conservation of
spin vector length is due to the spin precession with the tip moving over
the Bloch sphere. Note that there is no integral of motion corresponding to
the Runge-Lenz vector specific for the {Coulomb-like field in the 
Kepler problem without spin-orbit coupling.} Then, with the SOC included, the system 
having five coupled degrees of freedom, 
possesses only three integrals of motion. As a result, it exhibits an unusual
chaotic behavior as we shall see below.

Since the comprehensive description  of the Kepler trajectories can be done
with the $t=0$ initial conditions   
\begin{equation}
\mathbf{r}(0)=(r_0,0),\qquad \mathbf{p}(0)=(0,p_0),  \label{inc}
\end{equation}%
we employ these conditions for subsequent analysis. We will measure the 
energy in $e^{2}/r_{0}$ and momentum in $e\sqrt{m/r_{0}}$ units respectively. 
The condition for the system to be in a highly excited Rydberg state is $r_{0}\gg \hbar^{2}/me^{2}$.
In addition, in what follows we put $\hbar=m=e=1$.
In Eq. \eqref{inc},  $p_{0}^{2}/2-1\equiv E_{0}$ is the initial energy in the SOC absence, 
which determines the trajectory shape. In this case, the initial angular momentum $L(0)\equiv 
\sqrt{2\left( 1+E_{0}\right) }.$ Note that $E_{0}=-1/2$ determines the
special case of circular trajectory. Its special character will be demonstrated 
below.  The total conserved energy is $E=E_{0}-\alpha p_{0}\bar{\sigma}%
_{x}(0).$ We will vary the initial conditions by modifying the initial spin: 
$\bar{\bm\sigma }(0)=(\bar{\sigma}_{x}(0),\bar{\sigma}_{y}(0),\bar{\sigma}%
_{z}(0))$ and study the subsequent coupled spin-coordinate motion dependent
on the initial spin vector. {We shall demonstrate numerically that  
at sufficiently large $\alpha$ the chaotic trajectories emerge in the system under consideration.}

\section{Weak spin-orbit coupling: trajectory deformation}

\begin{figure}[t]
\begin{center}
\includegraphics*[width=1.0\columnwidth]{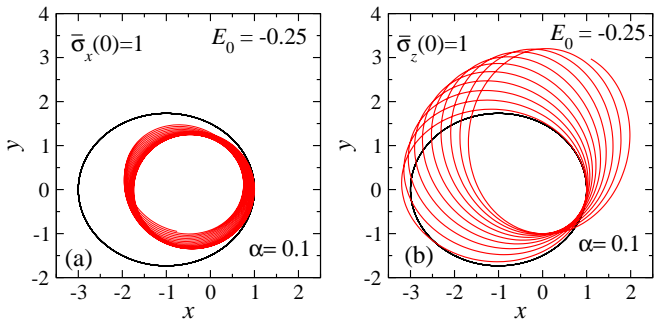}
\end{center}
\caption{ Trajectories for $E_{0}=-0.25$ and initial spins $\bar{%
 \sigma}_{x}(0)=1$ (a) and $\bar{ \sigma}_{z}(0) = 1$ (b). In
both panels $\alpha=0.1$. Elliptic line is the trajectory for $\alpha=0.$ }
\label{fig:trajectories}
\end{figure}

{We begin with the solution of equations of motion 
to describe the trajectories at weak SOC.} Here we can study
the dynamics by iterative procedure in $\alpha $ in Eqs. \eqref{dpdt},%
\eqref{dsdt} with the elliptic Kepler trajectory and time-independent spin
as zero-order approximation. In this case the anomalous velocity has two
main effects on the {trajectory shape}: (i) it shifts the position of the
point $r(T/2)$ (see Fig. \ref{fig:trajectory}) and (ii) it
rotates the entire trajectory (see Appendix A for details). 
Based on the total angular momentum and energy conservation,
we present the position of the particle at $t=T/2$ as: 
\begin{equation}
{r(T/2)}=\frac{L^{2}(T/2)}
{Z_{T/2}\mp\sqrt{Z_{T/2}^{2}+2EL^{2}(T/2)}}, 
\label{rn}
\end{equation}%
where $Z_{T/2}\equiv 1-\alpha L(T/2)\bar{\sigma}_{x}(T/2)$, $L(T/2)\equiv
L(0)+\Deltạ_{z},$ $2\Deltạ_{z}\equiv \bar{\sigma}_{z}(0)-\bar{\sigma}%
_{z}(T/2),$ and the upper (lower) sign corresponds to $E_{0}>-1/2$ ($E_{0}<-1/2
$). At $\alpha =0$ we have $r(T/2)=R$, where $R\equiv -(E_{0}+1)/E_{0}$ determines 
the limit of the trajectory in the absence of the SOC, and at
finite $\alpha$, we consider the shift $r(T/2)-R$. Analysis of Eq. %
\eqref{rn} shows that $E_{0}=-1/2$, where $1+2E_{0}L^{2}(0)=0,$ is the special
case (see Appendix A for details), which should be considered separately. To demonstrate the key role of the 
initial conditions, we consider two cases which are strongly different in terms of
the anomalous initial velocity.

We begin with $\bar{\sigma}_{x}(0)=1,$ where the initial velocity is $%
\mathbf{v}(0)=(0,p-\alpha )$. The first two iterations of Eqs. \eqref{dpdt}, \eqref{dsdt}
yield the leading terms in the spin components $\bar{\sigma}_{z}(T/2)=2\alpha \left( R+1\right) $
and $\bar{\sigma}_{x}(T/2)=1-2\alpha^{2}\left( R+1\right)^{2}$. It is seen
that the main contribution in Eq. \eqref{rn} comes from $\bar{\sigma}_{z}(T/2)$, which is
linear in $\alpha $, and we may safely disregard the $\sim\alpha^{2}$ correction in
$\bar{\sigma}_{x}(T/2).$ Then, at $\left\vert1+2E_{0}\right\vert \gg $ $\alpha ,$ 
the linear in $\alpha $ correction to
the position becomes: 
\begin{equation}
R-r(T/2)=-\frac{R}{l_{\mathrm{so}}}
\frac{1}{E_{0}}
\frac{\sqrt{2}}{\sqrt{1+E_{0}}}.  
\label{sx1}
\end{equation}
The initial condition $\bar{\sigma}_{z}(0)=1$ can be considered along the
same lines with $\mathbf{v}(0)=(0,p_{0})$ and $E=E_{0}$. We obtain two
main contributions to the shape of the trajectory: $\bar{\sigma}%
_{x}(T/2)=-2(R+1)/l_{\mathrm{so}}$ and $\bar{\sigma}%
_{z}(T/2)=1-2(R+1)^{2}/l_{\mathrm{so}}^{2},$ resulting at $\left\vert
1+2E_{0}\right\vert \gg $ $\alpha^{2}$ in  
\begin{equation}
R-r(T/2)=-\frac{R\left( R+1\right) }{l_{\mathrm{so}}^{2}}\frac{\sqrt{2}}{%
\sqrt{1+E_{0}}}.  \label{sz1}
\end{equation}
To illustrate the deformation of the trajectories, in Fig. \ref{fig:trajectories}, 
we present them at a moderate SOC for different initial
conditions. The shapes of the trajectories correspond well to Eqs. %
\eqref{sx1} and \eqref{sz1}.
At a sufficiently strong SOC, {corresponding to} a small $l_{\mathrm{so}},$ the {relevant}
quantities  such as deformations and rotations, become large, providing a hint to possible chaotic behavior.

\section{Chaos and order from disorder at strong coupling}

\begin{figure}[t]
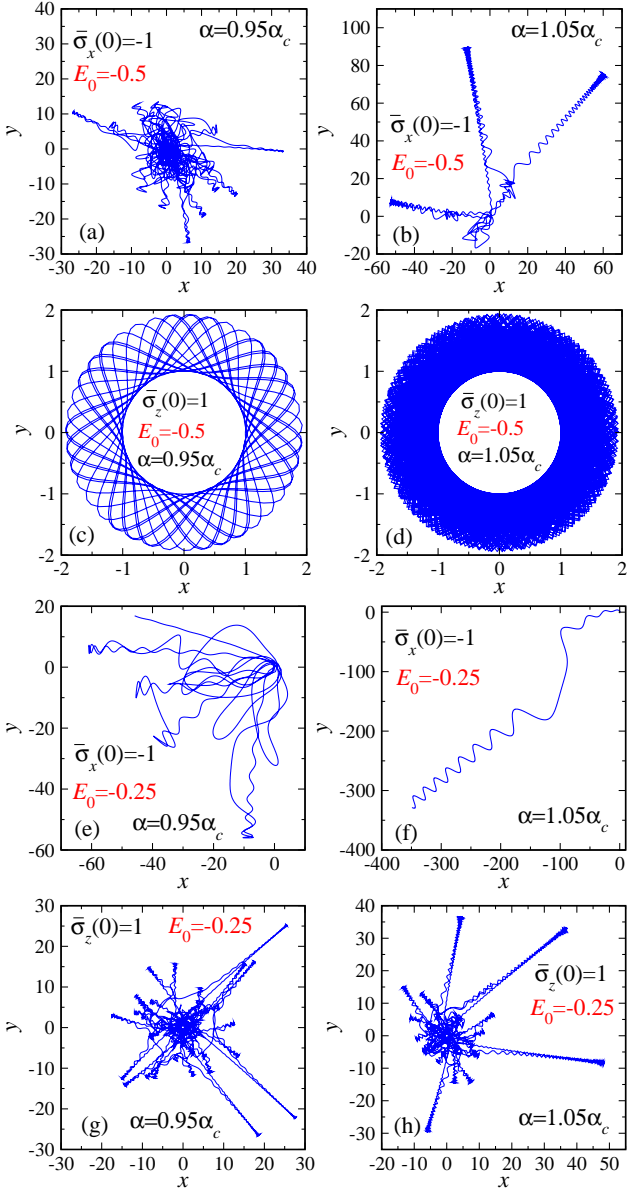

\begin{center}
\includegraphics*[ width=0.96\columnwidth]{figure3_upper.eps} \\[0pt]
\includegraphics*[ width=0.96\columnwidth]{figure3_lower.eps}
\end{center}
\caption{ The electron trajectories in the exciton with SOC for energies $%
E_{0}=-0.5$ ((a)-(d)) and $E_{0}=-0.25$ ((e)-(h)) for the parameters shown
in the panels. Left and right columns correspond to $ \alpha =0.95%
 \alpha_{c}$ and $ \alpha =1.05 \alpha_{c}$,
respectively. Here $ \alpha_{c}= \sqrt{2}(1- \sqrt{%
1+E_{0}})$ for $\bar{ \sigma}_{x}(0)=-1$ and $ \alpha_{c}=%
 \sqrt{-2E_{0}} $ for $\bar{ \sigma}_{z}(0)=1$. Panels (c) and
(d) demonstrate that these initial conditions belong to a domain of
stability with respect to SOC-induced chaotization. In all the panels except (c),
time $t<10^{3}$; in the panel (c) $t<2\cdot 10^{2}$ for a better resolution of the trajectory.}
\label{fig:randomization}
\end{figure}

Having discussed the effect of the relatively weak and moderate SOC, we can
establish a crossover from the weak to a strong coupling regime. The qualitative effects
of strong SOC appear when  $\alpha$ becomes of the order of the minimal Kepler velocity, $v_{\min}.$ 
For simplicity we consider the case $E_0 \to -0$.
In this limit $R=1/|E_0|$ and $L(0)=\sqrt{2},$ therefore, $v_{\min}=$ $\sqrt{2}|E_0|$ and the criterion of strong
SOC becomes $\alpha \gtrsim |E_0|$. However, there is a
subtle effect related to the role of the initial conditions - if the anomalous
velocity is initially zero, as in the $\bar{\sigma}_{z}(0)=1$ case, it needs some time to be developed. 
Taking into account that the spin precession yields $\bar{\sigma}_{x}(T/2)\sim\alpha R$, here the
condition of an immediate strong effect of SOC, becomes $\alpha^{2}R\sim |E_0|$. 

Now we can determine the critical value $\alpha_c$, which
permits ionization of the initial state and serves as a typical SOC value 
where the chaotic regime can be seen.  
The minimum of the SOC \eqref{hso} eigenenergy $E=-\alpha^{2}/2$ occurs at $p=\alpha$,
and the ionization is determined by the condition that the
total energy $E=E_0 -\alpha_{c}p_{0}\bar{\sigma}_{x}(0)$ equals {to} the SOC
eigenenergy minimum 
\begin{equation}  \label{wer}
E_0 -\alpha_c p_0\bar{\sigma}_{x}(0)=-\frac{\alpha_{c}^{2}}{2}.
\end{equation}
The solution of Eq. \eqref{wer} yields $\alpha_c=u_0+\sqrt{u_0^2-2E_0}$, where $%
u_0=p_{0}\bar{\sigma}_{x}(0),$ demonstrating a non-analytical dependence of 
$\alpha_{c}$ on the initial conditions. 

For $\bar{\sigma}_{x}(0)=0$ it is immediately seen that $\alpha_{c}=\sqrt{%
2\left\vert\,E_0 \right\vert }$, and this condition is stronger
than $\alpha \gtrsim |E_0|$. Therefore, even when the trajectory is strongly
modified by the Rashba coupling, the motion can still be finite. This
difference can be seen in the opposite limit as well: $E_0\to -1$ yields $%
p_0\to 0$ with $\alpha_c=\sqrt{2}\gg\,p_0.$ 

{We now address the possibility of occurrence of infinite trajectories, that is of the exciton ionization, 
at a given $\alpha$. To create such a trajectory (which is a consequence of the
initial bound state ionization),} one needs $\bar{\sigma}_{x}(0)<\bar{\sigma}_{x}^{\rm cr}(0)
\equiv\left(E_{0} +\alpha
^{2}/2\right) /\alpha p_{0}$. This equation determines the minimal value of $%
\alpha_{c},$ corresponding to $\bar{\sigma}_{x}(0)=-1:$ 
\begin{equation}
\alpha_{c}^{\min}=-p_{0}+\sqrt{p_{0}^{2}-2E_{0} } =\sqrt{2}\left(1-\sqrt{1+E_{0}}\right).
\label{acmin}
\end{equation}
At smaller $\alpha<\alpha_{c}^{\min}$ the motion is always finite and the SOC-induced
ionization is prohibited.

Therefore, if the total initial energy of the system approaches $-\alpha
^{2}/2$ from below, the particle shows a finite but long-range motion with the maximal 
distance $r_{\max}\sim\left(\alpha_{c}-\alpha\right)^{-1}$. In
contrast to the conventional Kepler problem, where the motion remains elliptic
when the negative total energy approaches zero, here it can become
chaotic. In general, condition $E=-\alpha^{2}/2$ determines a
multidimensional separatrix in the phase space augmented by spin subspace.
Note {that ionization condition} $E>-\alpha^{2}/2$ is a necessary, but not sufficient one - 
even when it is satisfied, the motion can still be finite, making a qualitative difference 
from the behavior without SOC. Indeed, at $\alpha=0,$ when $L$ is the integral 
of motion, the dynamics can be reduced to one-dimensional form with 
the effective potential energy $U_{\rm eff}(r)=-1/r + L^2/2r^2$ \cite{Landau} explicitly 
including $L$. This one-dimensional mapping results in 
$E_{0}>0$ being both the necessary and sufficient condition {for} the ionization. For 
nonzero $\alpha$ such a mapping cannot 
be done and the condition $E_{0}>-\alpha^{2}/2$ is not sufficient anymore.

\begin{figure}[t]
\begin{center}
\includegraphics*[width=1.07\columnwidth]{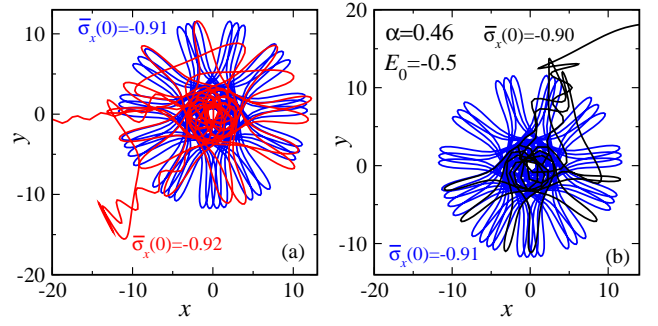} 
\end{center}
\caption{ Trajectories at different initial spins marked near the lines for
the parameters shown in the panel (b). Note that at $\bar{ \sigma}
_{x}(0)=-0.91$ the trajectory is flower-like (as presented for $t<2\times 10^{2}$). 
Here initial spin rotates in the $(xz)$
plane such that $\bar{ \sigma}_{y}(0)=0$.}
\label{fig:xaos}
\end{figure}

Now we provide evidence for the chaotic behavior based on numerical solutions of Eqs. \eqref{dpdt} and \eqref{dsdt}. 
The particle trajectories near the separatrix corresponding to the 
critical SOC are presented in Fig. \ref{fig:randomization}, demonstrating a variety of behaviors. The
minimal and maximal distance reached by the particle in this case are determined by the
conservation laws. Figure \ref{fig:randomization}, demonstrating all possible
behaviors of the SOC-augmented Kepler problem close to the critical coupling, can be considered as the main
result of our analysis. These behaviors include: 

1) strongly entangled chaotic trajectories with
"protuberances'' in Figs. \ref{fig:randomization}(a), (b), (e), (g), and (h), {as expected from the fact that 
the number of the integrals of motion is less than the number of the degrees of freedom}; 

2) {stable unperiodic orbits} filling regular ring-like areas in the $(xy)-$plane in Figs. \ref{fig:randomization}(c) and (d). 
This behavior resembles the trajectories in a system with a non-Coulomb potential $U(r)$ as  
presented, e.g., in Ref. \cite{Landau}, demonstrating a stability point in the Kepler problem with SOC, and 

3) relatively simple trajectories corresponding to the exciton ionization in Fig. \ref{fig:randomization}(f). 

\begin{figure}[t]
\begin{center}
\includegraphics*[width=0.85\columnwidth]{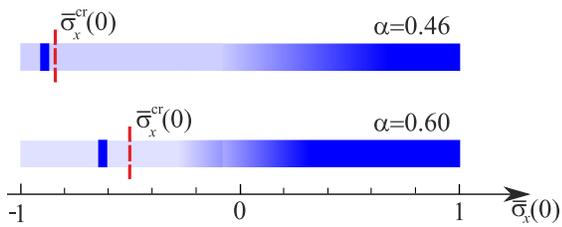}
\end{center}
\caption{ Different regimes of motion dependent on the 
initial spin component $\bar \sigma_x(0)$ exemplified by $xz$-plane rotation
for $\alpha=0.46$ and 0.6. In this case $\bar \sigma_y(0)$ = 0 and $\bar \sigma_z(0)=
\sqrt{1-\bar \sigma_x^2(0)}$. Dark (blue in color version) parts correspond
to a regular finite motion, while gray (light blue) parts correspond to chaotic one, either localized or delocalized. Thus,
the bars show a gradual development of the chaos-related
features following the regular motion. The necessary condition (boundary, dashed vertical lines) for ionization is given
by $\bar \sigma_z^{\rm{cr}}(0)$ defined above Eq. \eqref{acmin} with $\bar \sigma_z^{\rm{cr}}(0)=-0.86$ and 
$\bar \sigma_x^{\rm{cr}}(0)=-0.53$ for $\alpha=0.46$ and $\alpha=0.6$ respectively.
A narrow domain of a regular motion appears at $\bar \sigma_x(0)<\bar \sigma_x^{\rm{cr}}(0)$. 
These domains are determined by $\bar \sigma_x(0) \in (-0.915,-0.905)$ 
and $\bar \sigma_x(0) \in (-0.64,-0.6)$ for $\alpha=0.46$ and $\alpha=0.6$ respectively.}
\label{fig:regimes}
\end{figure}

{At large distances, $r\gg \pi l_{\mathrm{so}}$, the influence of Coulomb
potential on the velocity is negligible compared to that of the spin-orbit
coupling. Hence, in this case the electron motion is almost completely defined by 
SOC.} Latter fact permits analytical description of {both the trajectories
like} rare protuberances, where electron still returns to
the chaotic region with $r\lesssim \pi l_{\mathrm{so}},$ and ionization
{ones} with $r\to \infty.$  {Without loss of generality, here we concentrate on the case of} 
the particle characterized by weakly
time-dependent momentum, which we consider for definiteness to be constant in
time as $\mathbf{p}=p\left( \cos \theta ,\sin \theta \right) $. The total
energy now consists of two constant terms: the kinetic energy $p^{2}/2$ and the 
SOC-related contribution 
\begin{equation}\label{ukk1}
\epsilon_{\mathrm{so}}=\alpha \left( p_{x}\bar{\sigma}_{y}(0)-p_{y}\bar{%
\sigma}_{x}(0)\right).
\end{equation}
Hence, the precession of the SOC-induced spin component parallel to the momentum $\mathbf{p}\bar{\bm{\sigma}}$ reads: 
\begin{equation}
\mathbf{p}\bar{\bm{\sigma}}(\tilde{t})=
\mathbf{p}\bar{\bm{\sigma}}(0)\cos
\left( 2\alpha p\tilde{t}\right),\ \tilde{t}\equiv t-t_0,
\end{equation}%
where $t_{0}$ is the initial time for the constant-$p$ motion. By solving these
equations we obtain the velocities $v_{x}(\tilde{t})=p_{x}+\alpha \bar{\sigma%
}_{y}(\tilde{t})$ and $v_{y}(\tilde{t})=p_{y}-\alpha \bar{\sigma}_{x}(\tilde{%
t})$: 
\begin{eqnarray}
v_{x}(\tilde{t}) &=&p_{x}+\alpha \left[ \bar{\sigma}_{y}(\widetilde{0})f_{1}(%
\tilde{t})-\bar{\sigma}_{x}(\widetilde{0})f_{2}(\tilde{t})\right] ,  \notag
\\
v_{y}(\tilde{t}) &=&p_{y}-\alpha \left[ \bar{\sigma}_{x}(\widetilde{0})f_{3}(%
\tilde{t})-\bar{\sigma}_{y}(\widetilde{0})f_{2}(\tilde{t})\right] ,
\label{sky}
\end{eqnarray}%
where $f_{1}(t)=1$\thinspace $-2\sin^{2}\theta \sin^{2}\phi $, $%
f_{2}(t)=\sin 2\theta \sin^{2}\phi $, and $f_{3}(t)=1-2\cos^{2}\theta \sin
^{2}\phi $ with $\phi \equiv \,\alpha pt.$ As we can see from Eqs.\eqref{sky},
the motion has an oscillatory character corresponding to Fig. \ref%
{fig:randomization}. 

Equations \eqref{sky} have an interesting limit. Consider particle
with momentum $p=\alpha.$
Then, the energy $E=\alpha^{2}/2+\alpha^{2}\left( \cos \theta \bar{\sigma}
_{y}(0)-\sin \theta \bar{\sigma}_{x}(0)\right) $ has the minimum $E=-\alpha^{2}/2
$ at $\bar{\sigma}_{y}(0)=-\cos \theta ,$ $\bar{\sigma}_{x}(0)=\sin \theta .$ Slightly
away from the minimum by taking: $\bar{\sigma}_{y}(0)=-\cos \left( \theta +\eta
\right) ,$ $\bar{\sigma}̣_{x}(0)=\sin \left( \theta +\eta \right) $ with small $%
\eta\ll\,1$ we obtain $v_{x}=\alpha \eta \sin \theta \cos \left( 2\alpha
^{2}t\right) $ and $v_{y}=-\alpha \eta \cos \theta \cos \left( 2\alpha
^{2}t\right).$ Here the
nonzero velocity appears only due to deviation of the energy from $-\alpha
^{2}/2$ with $E=-\alpha^{2}\left(1-\eta^{2}\right)/2.$ This corresponds to small but fast oscillations 
in the "protuberances" in Fig. \ref{fig:randomization}. These
oscillations are an extreme manifestation of a \textit{Zitterbewegung} \cite{Schliemann,Winkler} 
typical for spin-orbit coupled systems. 

{An important characteristic feature of a chaotic motion such as 
strong dependence on the initial conditions is presented in Fig. \ref{fig:xaos}.} It clearly demonstrates a
transition from {chaotic to {confined} regular} trajectory (Fig. \ref{fig:xaos}(a)) in the
stability domain and {\emph{vice versa}} (Fig. \ref{fig:xaos}(b)) at small variations in the
initial spin direction. The stability point here is approximately  $\bar{\sigma}_{x}(0)=-0.91$.
Note that the chaotic trajectories reported in Fig. \ref{fig:xaos}(a) and (b) (corresponding to $\bar{\sigma}_{x}(0)=-0.9$ 
and 0.92 respectively) are qualitatively similar to those in Fig. \ref{fig:randomization}(b) since the value of $\alpha$
is larger than the critical $\alpha_{c}$ obtained with Eq. \eqref{wer} for both initial spin orientations. 

In general, the {trajectory shape depends on the SOC} strength and the 
entire set of the initial conditions, making the analysis very cumbersome. 
To obtain a semi-quantitative pattern of different regimes of the motion, 
the domain of initial conditions in  
Fig. \ref{fig:xaos} can be extended to the {entire $\bar{\sigma}_{x}(0)=(-1,1)$ domain}. 
The results are presented in Fig. \ref{fig:regimes} for two values of the constant $\alpha$. {This
Figure shows a gradual development of chaos-related features such as increasing entanglement of 
the trajectories and length of the protuberances following the regular
behavior and can be regarded as a "stability diagram" of the system under consideration.} Another important feature, which already 
follows from Fig. \ref{fig:xaos}, is a reentrance
of regular trajectories from chaotic ones and {\em{vice versa}} 
near the threshold values $\bar{\sigma}_{x}^{\rm cr}(0)$. 
Latter effect can be considered as "order from disorder" driven by  variation in the 
initial spin direction.

\section{Conclusions}

We have studied the emergence of chaos in the classical Kepler
problem with the Rashba spin-orbit coupling. Such systems can be experimentally realized, for instance, 
in highly excited Rydberg-like states of two-dimensional excitons in semiconductor structures. 
At a weak Rashba coupling, corresponding to small interaction constants $\alpha$, the elliptic 
Kepler trajectories are modified in shape and orientation 
and densely fill a part of the particle motion plane.
This behavior is somewhat similar, although not identical, to that of the Kepler problem 
trajectories in the potentials 
$-1/r^{b}$, where $b>0$ and $b\ne 1$ \cite{Landau}.
With the increase in $\alpha$, the deformations of the trajectories become strong and in the vicinity 
of the critical $\alpha_{c}$ corresponding to the exciton breakdown, the 
trajectories can become chaotic. {Typical chaotic} trajectory can be described as a highly entangled 
path in the vicinity of the initial position with rare long
protuberances increasing in length at approaching the 
ionization threshold. At a sufficiently large $\alpha,$ the SOC assisted
ionization becomes possible and the bound state disappears.  
The reason for the chaos lies in the fact that the system lost integrability since 
in possesses only to integrals of motion for its four degrees of freedom. Dynamically, 
this effect is clearly seen in the equations of motion including the anomalous spin-dependent velocity term. 

As typical for a chaotic system, its dynamics exhibits a critical dependence on the initial conditions, 
demonstrating the transition from a chaotic to a regular behavior 
and \emph{vice versa} caused by  very small
variations in the initial spin orientation. The latter feature could be referred to as 
reentrant "order-from-disorder" transition. Our results can have important
implications for the properties of  semiclassical Rydberg-like 
states of excitons in two-dimensional structures, where a chaotic regime due 
to spin-orbit coupling may be developed. 

To make relation to possible observations, we mention that the typical energies of highly
excited Rydberg 2D excitons in semiconductor quantum wells are of the order of 0.1 meV and the size is around 100 nm \cite{Excitons}. 
Therefore, a typical length of the chaotic protuberances is about $10^3$ nm (1 $\mu{\rm m}$)
and a typical time the electron 
spends in a long protuberance is about a nanosecond. We mention here several realizations, where 
predicted chaotic trajectories can play a role. 
For example, they can be relevant for interaction between distant 2D excitons. Indeed, 
the long-range chaotic trajectories presented in Fig. \ref{fig:randomization} are 
characterized by large dipole moments $e{\mathbf r}(t)$,
enhancing long-range electric fields of the excitons. The chaotic trajectories of 
electrons from different excitons can interlace, which alters the exciton-exciton interaction leading either 
to their ionization or to a modified lateral motion. In addition, this chaotic behavior can strongly influence 
relaxation of electron energy in the Rydberg excitons by emission of phonons. 
Instead of a process with well-defined time dependence, the energy relaxation from a highly excited to the 
ground state may become chaotic and largely unpredictable. Moreover, the electrons at those remote trajectories, 
can eventually be trapped in distant electrostatic fields (e.g. Ref. \cite{Butovtraps}) or at the system boundaries.

\begin{acknowledgments}
E.Y.S. acknowledges the support by the Spanish Ministry of Economy, 
Industry, and Competitiveness and the European Regional 
Development Fund FEDER through Grant No. FIS2015-67161-P (MINECO/FEDER), 
and Grupos Consolidados UPV/EHU del Gobierno Vasco (IT-986-16).
\end{acknowledgments}

\appendix

\section{Trajectory shape at weak spin-orbit coupling}
\begin{widetext}
Based on the energy and angular momentum conservation, we calculate the particle's position
at $t=T/2.$ At this time instant, the energy can be written as:
\begin{equation}
E=\frac{p^{2}\left( T/2\right) }{2}+\alpha p\left( T/2\right) \bar{\sigma}
_{x}(T/2)-\frac{1}{r\left( T/2\right)},\ p(T/2) =\frac{L(T/2)}{r(T/2)}.
\end{equation}%
The total angular momentum conservation yields $L\left(T/2\right) =L(0)+\left( \bar{\sigma}
_{z}(0)-\bar{\sigma}_{z}(T/2)\right)/2.$ Thus, we obtain: 
\begin{equation}\label{kuka1}
\frac{\left(L(0)+\Delta_{z}\right)^{2}}{r^{2}\left( T/2\right) }+\frac{2}{%
r\left( T/2\right) }\left( \alpha \left(L(0)+\Delta_{z}\right) \sigma_{x}(T/2)-1\right) -2E=0,
\end{equation}%
where $\Delta_{z}\equiv\left(\bar{\sigma}_{z}(0)-\bar{\sigma}_{z}(T/2)\right)/2$.
The solution of \eqref{kuka1} with respect to $r(T/2)$ yields
\begin{equation}
r\left(T/2\right)=\frac
{\left(L(0)+\Delta_{z}\right)^{2}}
{
1-\alpha\left(L(0)+\Delta_{z}\right)\bar{\sigma}_{x}(T/2)
\mp 
\sqrt{(\alpha(L(0)+\Delta_{z})\bar{\sigma}_{x}(T/2)-1)^{2}+2E\left(L(0)+\Delta_{z}\right)^{2}}
}
,
\label{1rT} 
\end{equation}%
where upper (lower) sign corresponds to $E_{0}>-1/2$ ($E_{0}<-1/2$). This equation is tantamount 
to Eq. \eqref{rn} of the main text. 

{Now we apply} Eq. \eqref{1rT} to the initial conditions  $\bar{\sigma}_{x}(0)=1$, 
where $E=E_{0}-\alpha \sqrt{2\left( 1+E_{0}\right)},$ and 
$L(0)=\sqrt{2\left(1+E_{0}\right)}.$ The equations for spin precession with respect to
anomalous velocity read: 
\begin{equation}
\dot{\bar{\sigma}}_{x} =2\alpha \left( v_{x}-\alpha {\bar{\sigma}}_{y}\right)
{\bar{\sigma}}_{z};\qquad  
\dot{\bar{\sigma}}_{y}=-2\alpha \left( v_{y}+\alpha {\bar{\sigma}}_{x}\right)
\bar{\sigma}_{z};\qquad 
\dot{\bar{\sigma}}_{z} =-2\alpha \left( v_{x}{\bar{\sigma}}_{x}+v_{y}{\bar{\sigma}}
_{y}\right). 
\end{equation}%

We begin with the time dependence of $\bar{\sigma}_{z}(t)$ and obtain%
\begin{equation}
\bar{\sigma}_{z}(t)=-2\alpha \left( x(t)-x(0)\right) 
\equiv-2\alpha \left(x(t)-1\right), 
\end{equation}%
resulting in $\bar{\sigma}_{z}(T/2)=2\alpha \left( R+1\right) $ and 
$L(T/2)=L(0)-\sigma_{z}(T/2)/{2}=L-\alpha\left(R+1\right).$ 
Second iteration yields $%
\bar{\sigma}_{x}(T/2)=1-2\left(\alpha\left( R+1\right) \right)^{2},$ 
providing the contribution which is not important for the linear in 
$\alpha$ approximation. Assuming $E_{0}>-1/2$, for this case we obtain%
\begin{equation}
{r\left(T/2\right)}=\frac 
{\left(L(0)-\alpha\left(R+1\right)\right)^{2}}
{
1-\alpha\left(L(0)-\alpha\left(R+1\right)\right)
-\sqrt{\left(\alpha\left(L(0)-\alpha \left(R+1\right)\right)-1\right)^{2}+
2E\left(L(0)-\alpha\left(R+1\right)\right)^2}
}
.
\label{rnS}
\end{equation}%
Equation \eqref{sx1} of the main text immediately follows from the above Eq. \eqref{rnS}. 

Now we {express} $\left(\alpha\left(L(0)-\alpha \left(
R+1\right) \right) -1\right) ^{2}+2E\left(L(0)\,-\alpha \left( R+1\right) %
\right)^{2}$ in terms of the energy $E_{0}$ and obtain
\begin{eqnarray}
&&\left( \alpha \left(L(0)-\alpha \left( R+1\right)\right)-1\right) ^{2}+2E%
\left(L(0)-\alpha \left( R+1\right)\right)^{2}= \\
&=&\left( 1+2E_{0}\right) ^{2}-2\alpha L(0)\left( 1+2E_{0}\right) +\alpha
^{2}\left( \sqrt{2\left( 1+E_{0}\right) }-8\frac{1+E_{0}}{E_{0}}\right)+{\cal O}(\alpha^3). \nonumber
\end{eqnarray}%
Therefore, at $1+2E_{0}=0$, the corresponding circular trajectory 
cannot be treated perturbatively and the condition 
$\left|1+2E_{0}\right|\gg \alpha$ is required for the applicability of the 
perturbation theory at given $\alpha\ll\,1.$

\section{Trajectories at $E_{0}<-1/2$.}

\begin{figure}
\begin{center}
\includegraphics*[width=14cm]{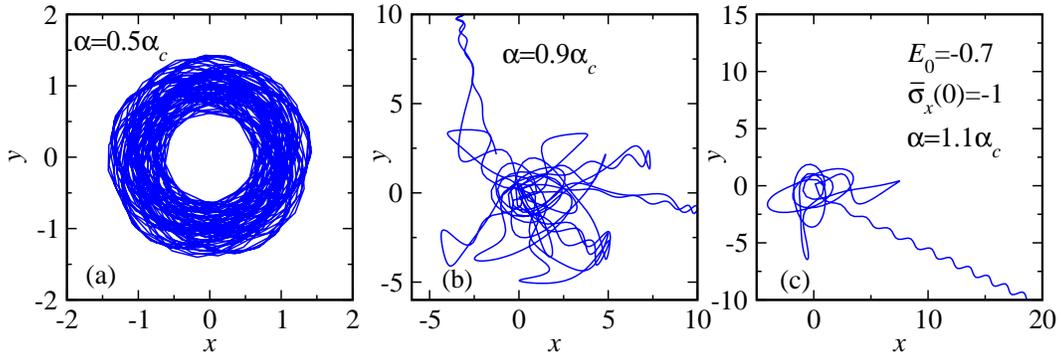}
\end{center}
\caption{ Particle trajectory for different values of SOC constant as shown in the 
panels. (a) Moderate $\alpha$ with the trajectory filling the allowed part 
of the $(xy)$ plane, (b) a chaotic trajectory
at $\alpha$ close to but smaller than the critical value, and (c) ionization 
trajectory at $\alpha>\alpha_{c}$. 
In all panels $E_{0}=-0.7$ and initial spin $\bar{\sigma}_{x}(0)=-1.$}
\label{fig:suppl}
\end{figure}

In the main text, we have considered three principal realizations of trajectories with $E_{0}\geq\,-1/2$ 
as shown in Fig.\ref{fig:randomization} there. 
Here we {complement} the corresponding analysis by typical results for $E_0<-1/2$
at different values of the SOC constant.
Figure \ref{fig:suppl} shows that the behavior of 
trajectories corresponds to the chaos
emergence near the critical $\alpha_{c}=\sqrt{2}(1-\sqrt{1+E_{0}})$,
followed by exciton ionization with $\alpha$ increase. This result follows  
from the general analysis in the main text. 
\end{widetext}


\begin{thebibliography}{99}
\bibitem{alpha} Although $\alpha$ can be either positive or negative, we
consider here only $\alpha>0$ realization.
\end{thebibliography}
\end{document}